\documentclass[]{spie}  

\usepackage{amsmath,amsfonts,amssymb}
\usepackage{graphicx}
\usepackage[colorlinks=true, allcolors=blue]{hyperref}

\title{Measuring the linear gain performance for the OrcaQuest CMOS imaging sensor}

\author[a]{Zain Azam}
\author[b]{Aditya Khandelwal}
\author[a]{Ryan Dungee}
\author[c]{Ting S. Li}
\author[a]{Sarik Jeram}
\author[a]{Shaojie Chen}
\affil[a]{Dunlap Institute for Astronomy and Astrophysics, University of Toronto, 50 St. George
Street, Toronto, ON, Canada M5S 3H4}
\affil[b]{Steward Observatory, University of Arizona, 933 N. Cherry Avenue, Tucson, Arizona, 85721, USA}
\affil[c]{University of Toronto, 27 King's College Circle, Toronto, Ontario, M5S 1A1, Canada}

\authorinfo{Further author information: (Send correspondence to Z.Y.A.)\\Z.Y.A.: E-mail: zain.azam@mail.utoronto.ca}

\begin{document} 
\maketitle

\begin{abstract}
The OrcaQuest CMOS sensor by Hamamatsu Photonics has been found to exhibit a non-linear gain curve at low exposure times. This effect is characterized through the various exposure tests and data reduction pipelines detailed in this report, and a more accurate gain characterization curve is provided. Such data can be used to provide more consistent measures of pixel gain, linearity, and high framerate images.
\end{abstract}

\section{INTRODUCTION}
\label{sec:intro}  
The two main types of imaging sensors used within both science and industry are Complementary Metal Oxide Semiconductors (CMOS) and Charge Coupled Devices (CCDs). Historically, CCD cameras have dominated scientific markets. They did suffer from many reliability issues, as their line-by-line readout method led to defects and noise patterns which could potentially affect entire rows of pixels, instead of just one, as in CMOS. However, their superior science properties maintained their place in scientific markets, while CMOS continued to grow in the consumer technology sector.

In recent years however, CMOS sensors have been known to outpace CCD sensors in various properties favorable for astronomical imaging, including high linearity, high quantum efficiency, and efficient global shutters over rolling shutters, making CCDs redundant for most applications. Much of this has occurred due to high investments in the consumer CMOS sensor market \cite{Khandelwal_2024}. These cameras are distinguished from consumer cameras by the term scientific CMOS (sCMOS), and trade various properties relative to consumers such as compression, visual appeal, and reliability for science properties such as low noise levels -- often only a few electrons or sub-electron -- and extremely high dynamic range, meaning that both low and high signal brightnesses can be captured on a single sensor \cite{Teledyne}.

The Hamamatsu OrcaQuest is an example of one of these sCMOS sensors. Through its alternative readout methods such as UltraQuiet (UQ) -- which reduces the camera  framerate for increased sensitivity -- or Photon Number Resolution (PNR) -- which directly measures photon counts without any additional data conversion -- the OrcaQuest is able to resolve images at extremely low light levels with significant accuracy down to only a few electrons per pixel \cite{Hamamatsu-Photonics}. Features of this camera include sub-electron read noise of $0.43e^{-}$, a $4096 \times 2304$ pixel sensor area, and a high dynamic range supporting up to 16 bits at $0.107$ ADU per electron \cite{Khandelwal_2024}.

One important measure of these sensors for astronomical applications is linearity, which describes the accuracy of how signal level follows exposure time. Linearity is imperative for any application that requires high contrast, and accurate relative brightness between pixels, such as photometry and spectroscopy -- as these applications will make use of the entire signal range of the camera. The OrcaQuest has been confirmed to exhibit linearity exceeding 99\% \cite{Khandelwal_2024}, but at specific, low signal ranges, this amount can deviate significantly. This is of particular interest because many estimates of linearity are calculated by averaging the linearity over linearly spaced signal levels. This type of calculation mixes highly nonlinear signal regimes with the highly linear signal regimes indiscriminately, and may not provide an accurate measure depending on the signal level used. As this report will demonstrate, linearity on the OrcaQuest tends to have localized trends in the low exposure region which may be overlooked by a typical linearity calculation.

In this report, we investigate how the linearity of the device deviates significantly for low light levels and present a method to replicate this characterization for other sensors. In section two, we describe how this model was derived, the results for our specific sensor, and the method used to perform this calibration for any other sensor. In section three, we evaluate the performance gains and accuracy of this calibration.

\section{Methods}
\label{sec:title}
\subsection{Data collection}
Data was collected from a 590nm light source at logarithmically spaced exposure times, from the minimum possible exposure time of $7.2\times10^{-6}$ seconds to a signal level just under the pixel saturation limit of 65535 ADU. These datasets were collected at mean pixel rates of 15 000 ADU/s, 20 000 ADU/s, and 29 000 ADU/s in order to eliminate or identify biases that may be related to light source brightness. Additionally, data was collected under both the standard (STD) and UltraQuiet (UQ) modes, two readout modes which are supported by the OrcaQuest. All three datasets for each readout mode were collated into a single dataset which was used for curve fitting. The remainder of data reduction occurs in logarithmic space.

\subsection{Data cleaning}
Bias frames and dark frames were removed from each image. Each bias and dark were collected at the beginning of the experiment at -20C through air cooling.

In order to remove the quantization effect at low exposure levels, a Savitzky-Golay filter with a window size of 2000 samples is employed to smooth data at low signal levels. Alternative methods of removing quantization such as downsampling and binning were experimented with, but but were noisy or did not represent the trend of the data under quantization. It is important to remove the smoothing at higher exposures in order to avoid filtering out the feature at 20 000 ADU by oversmoothing. To achieve this, we use a sigmoidal weighting of the form $\frac{1}{1 -10 (t - 0.5)}$, where $t$ is the fraction of logarithmically spaced samples up to and including this sample.

   \begin{figure} [htbp]
   \begin{center}
   \includegraphics[height=10cm]{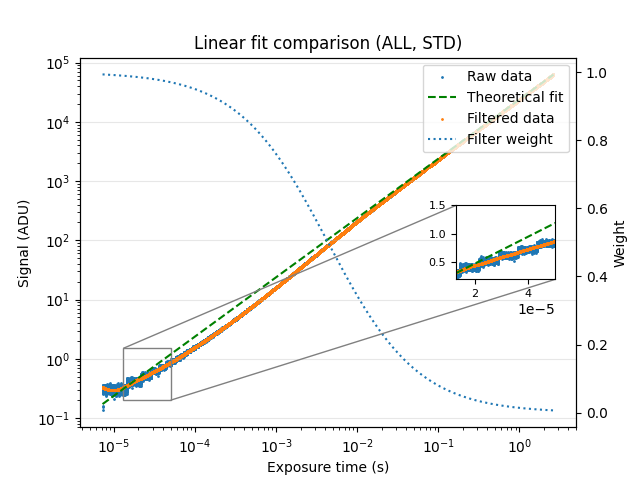}
	\end{center}
   \caption[example] 
   { \label{fig:linallstd} Linear fit, global signal levels by exposure, and smoothed data for standard (STD) mode. "Filter weight" shows the strength of curve smoothing. Raw data shown represents all datasets together after exposure time normalization. UltraQuiet (UQ) readout is shown in Appendix \ref{sec:uq}.}
   \end{figure}

Already in Figure \ref{fig:linallstd}, it is possible to see the curve undershoot the theoretical model for the earlier half of the data. A clearer numerical model of this is shown in the next section.

\subsection{Residual model}

The ADU rate of each dataset was calculated by taking 100 frames at a high signal level (25 000 to 65 000 ADU) and dividing their average ADU count by the exposure time, which provides the dataset's expected ADU rate. Using a high signal level to calculate the reference for ADU rate is important to avoid nonlinearities that occur at lower exposure levels.
\begin{equation}
    \text{ADU rate} = \text{Mean signal level at reference} \times \text{Exposure time}
\end{equation}
These samples are then normalized to a reference fit of 25 000 ADU/s.
\begin{equation}
    \text{Normalized exposure time} = \text{Exposure time} \times \frac{\text{25 000}}{\text{ADU rate}}
\end{equation}

Finally, the percentage deviations from the idealized linear estimate of 25 000 ADU counts per second were calculated for each sample by dividing the estimate by the smoothed sample values, normalized by brightness.

\begin{equation}
    \text{Idealized signal level} = \text{25 000} \times \text{Exposure time}
\end{equation}

\begin{figure} [htbp]
   \begin{center}
   \includegraphics[height=10cm]{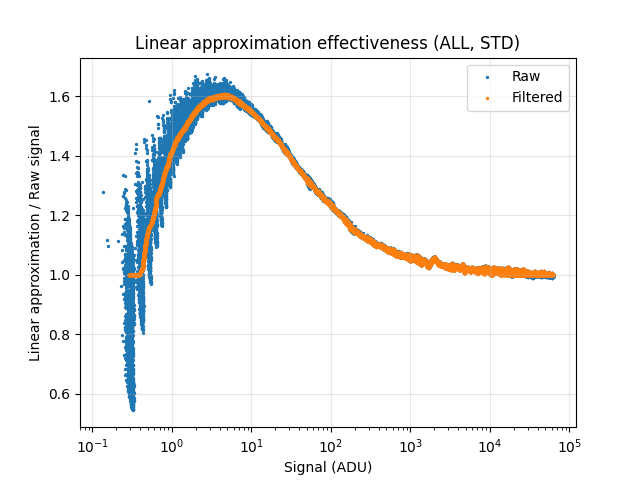}
	\end{center}
   \caption[example] 
   { \label{fig:rawscales} Raw versus smoothed residuals for STD readout mode. UQ readout is shown in Appendix \ref{sec:uq}.}
   \end{figure}

There is a high amount of scattering early due to quantization effects, which are mostly remedied by the smoothing filter. The degree of smoothing may slightly affect the final curve, so it is important to plot both when performing this calibration to verify whether the raw data is accurately represented by the smooth data such that it passes near the median of each quantization level. For values of the lowest quantization (less than 0.107 ADU), we cap the residual at 1.0. A reference plot can be seen in Figure \ref{fig:rawscales}.

\subsection{Readout mode fluctuations}
\begin{figure} [htbp]
   \begin{center}
   \begin{tabular}{c c} 
   \includegraphics[height=6cm]{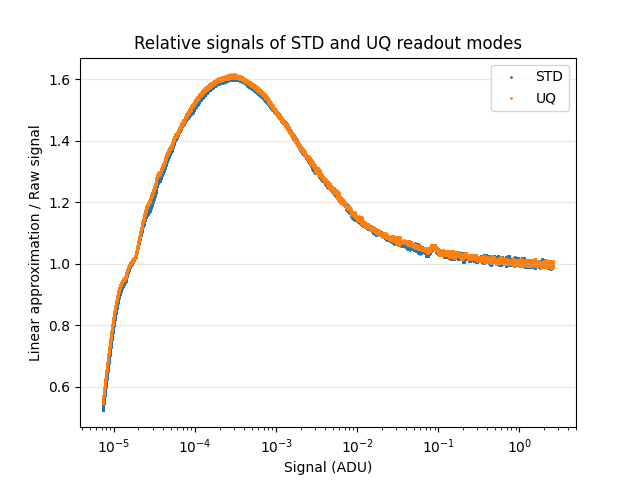} & \includegraphics[height=6cm]{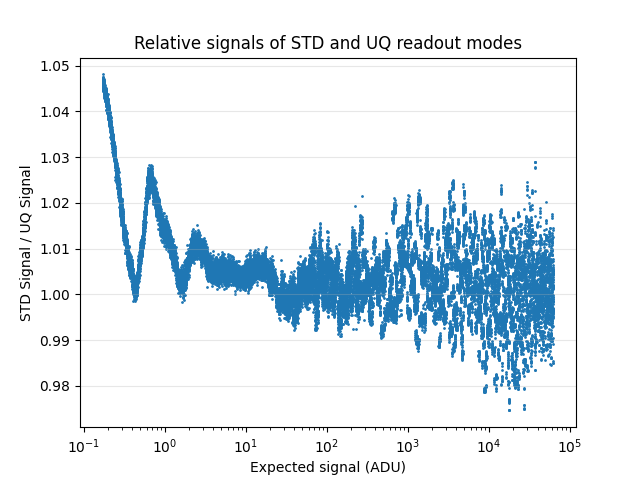}
	\end{tabular}
	\end{center}
   \caption[example] 
   { \label{fig:scales} Left: Percentage deviation from linear estimate for both STD and UQ readout. Right: Ratio between STD readout and UQ readout versus exposure time. There is minor additional scattering due to inter-dataset variations.}
   \end{figure}

We also make a comparison between the distribution for STD and UQ mode; UQ shows slightly higher deviations for very low exposures but remains within 5\% of standard readout as visible in Figure \ref{fig:scales}. Since there is a small deviation, it is recommended to perform this calibration process for each separate readout mode on future sensors.

\subsection{Dataset fluctuations}
\begin{figure} [ht]
   \begin{center}
   \begin{tabular}{c c} 
   \includegraphics[height=10cm]{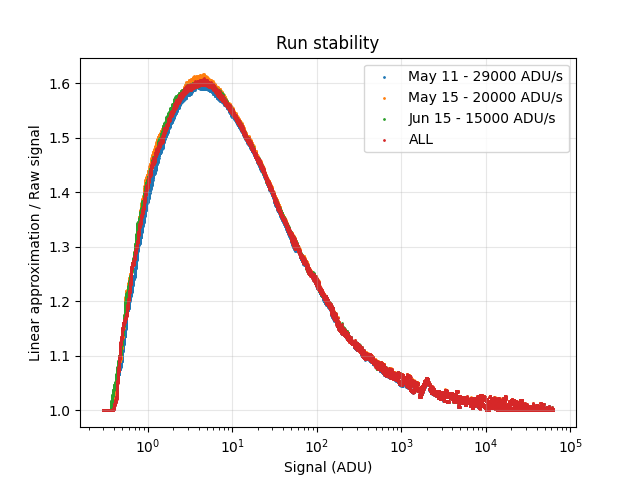}
	\end{tabular}
	\end{center}
   \caption[example] 
   { \label{fig:scatter} Dataset scattering for STD readout mode.}
   \end{figure}

Finally, we recognize a slight scattering that occurs between datasets of different brightnesses. The curves in Figure \ref{fig:scatter} are generated by employing the same data processing method described earlier at a lower filter window size of 1200 samples, but for each individual dataset, while the red curve employs all datapoints at once. Out of these datasets, the variation peaks at a 2\% difference between the May 11 and May 15 datasets near 5 ADU (46-47 $e^{-}$). Although small, it is unclear whether this is related to systematic differences during the data collection process, or features of the sensor, so it is recommended to use multiple datasets for this calibration.

\subsection{Data fitting}
Data is then placed into a cubic spline-fitting tool partitioned into 5 intervals, using a least-squares cubic regression over each interval. The boundaries for each interval were further tuned by optimizing for least-squares over the entire curve using numerical minimization techniques. Suggested initial values for this computation are provided in Appendix \ref{sec:knots}. Notably, three knots are used at the $2.32\times10^3$ ADU boundary, creating a derivative discontinuity to properly model the jump. In order to create a completely smooth model, the number of knots may be increased, but will involve a more complex fit which is not described in this report. Endpoints are also pegged to fall within $1\%$ of $1.0$ by the optimizer, to ensure continuity of the curve between the perfectly linear (constant value residual) and nonlinear regions.

The values for each of these curves are provided in Table \ref{tab:curvesstd} and \ref{tab:curvesuq}. Values outside of this range use no correction.
   
   \begin{table}[htbp]
\caption{Correction factor approximation for this study's OrcaQuest (STD)} {\label{tab:curvesstd}}
\label{tab:Multimedia-Specifications}
\begin{center}
\begin{tabular}{|l|l|l|}
\hline
\rule[-1ex]{0pt}{3.5ex}  Label & Signal range (ADU) & Linearity correction factor (unitless)  \\
\hline
\rule[-1ex]{0pt}{3.5ex}  A & $3.79\times10^{-1}<x<6.89\times10^{1}$ & $0.143x^3 -0.849x^2 + 1.306x + 1.000$   \\
\hline
\rule[-1ex]{0pt}{3.5ex}  B & $6.89\times10^{1}<x<1.82\times10^{3}$ & $-0.061x^3 + 0.189x^2 - 0.288x + 1.222$  \\
\hline
\rule[-1ex]{0pt}{3.5ex}  C & $1.82\times10^{3}<x<2.32\times10^{3}$ & $9.993x^3 -4.371x^2 + 0.440x + 1.037$  \\
\hline
\rule[-1ex]{0pt}{3.5ex}  D & $2.32\times10^{3}<x<3.05\times10^{3}$ & $-10.875x^3 + 1.625x^2 -0.109x + 1.030$  \\
\hline
\rule[-1ex]{0pt}{3.5ex}  E & $3.05\times10^{3}<x<4.79\times10^{4}$ & $-0.001x^3 -0.007x^2 -0.028x + 1.027$  \\
\hline
\end{tabular}
\end{center}
\end{table}

   \begin{table}[htbp]
\caption{Correction factor approximation for this study's OrcaQuest (UQ)} {\label{tab:curvesuq}}
\label{tab:Multimedia-Specifications}
\begin{center}
\begin{tabular}{|l|l|l|}
\hline
\rule[-1ex]{0pt}{3.5ex}  Label & Signal range (ADU) & Linearity correction factor (unitless)  \\
\hline
\rule[-1ex]{0pt}{3.5ex}  A & $2.83\times10^{-1}<x<1.00\times10^{2}$ & $0.114x^3 -0.734x^2 + 1.220x + 1.000$   \\
\hline
\rule[-1ex]{0pt}{3.5ex}  B & $1.00\times10^{1}<x<1.78\times10^{3}$ & $-0.015x^3 + 0.137x^2 -0.301x + 1.228$  \\
\hline
\rule[-1ex]{0pt}{3.5ex}  C & $1.78\times10^{3}<x<2.82\times10^{3}$ & $8.683x^3 -3.693x^2 + 0.358x + 1.037$  \\
\hline
\rule[-1ex]{0pt}{3.5ex}  D & $2.82\times10^{3}<x<3.16\times10^{3}$ & $-10.721x^3 + 1.517x^2 -0.077x + 1.030$  \\
\hline
\rule[-1ex]{0pt}{3.5ex}  E & $3.16\times10^{3}<x<4.78\times10^{4}$ & $0.069x^3 -0.091x^2 -0.006x + 1.029$  \\
\hline
\end{tabular}
\end{center}
\end{table}

   \begin{figure} [htbp]
   \begin{center}
   \includegraphics[height=10cm]{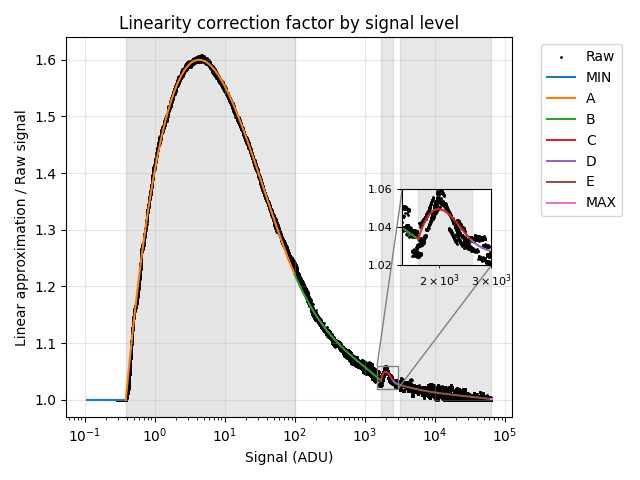} \\
   \includegraphics[height=10cm]{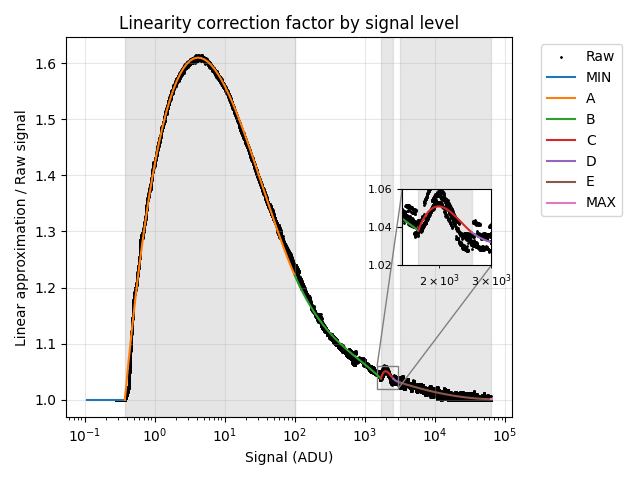}
	\end{center}
   \caption[example] 
   { \label{fig:correct} Correction curve for both STD (top) and UQ (bottom) readout modes.}
   \end{figure}

Residuals were also measured in another OrcaQuest \cite{Strakhov_2023}, which exhibits similar characteristics -- a peak residual at very small exposures, and a bump, likely related to a change in ADC for readout, at 20 000 ADU. However, the peak value of these residuals is 1.6 for this detector and 1.4 for the other detector. It is unclear whether this is related to the way the curve is calibrated, the way the linear fit is produced, or a detector-to-detector variation. For this reason, we recommend reproducing the steps in this report for any detector in order to eliminate any possible detector-to-detector variations. The fits for this particular camera can be seen in Figure \ref{fig:correct}.

\newpage
\section{Evaluation of Performance}
In order to evaluate the model, we compare this against a new dataset at an ADU rate of 15 000 ADU/s up to 1 second. In order to apply the model, we simply multiply each sample in the dataset by the corresponding factor based on its current intensity. We investigate how closely the residuals move closer to 1. We also see if the fit is more or less effective under certain signal powers.

\begin{figure} [ht]
   \begin{center}
   \begin{tabular}{c c} 
   \includegraphics[height=8cm]{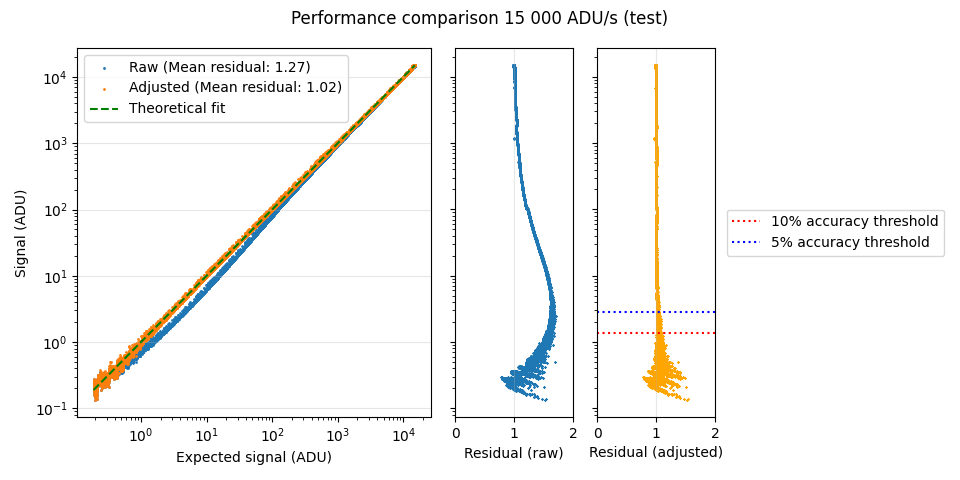}
	\end{tabular}
	\end{center}
   \caption[example] 
   { \label{fig:fix} Comparison of residuals for raw and adjusted data.}
   \end{figure}

In Figure \ref{fig:fix}, the residuals remain high under low exposures due to quantization, but this is to be expected -- the fit does not account for this. This is a marked increase from the peak deviation of 60\% when using a naive linear approximation. The mean residual value improves to 2\% which is very close to the intended value of 0\% and a strong improvement from the previous 27\%. There is also a strong improvement in its standard deviation which falls fivefold, meaning most samples converge around 1\%.

Additionally, we provide accuracy thresholds for 5\% and 10\% under this dataset. The approximation corrects the residuals to values below 10\% at 1 ADU, and to values below 5\% at 3 ADU. On the other hand, while uncorrected, the 10\% threshold occurs at 200 ADU. This proves that the model shifts low level exposures to a much more linear trend -- in this case the expected rate is 15 000 ADU/s, but the same correction can be applied over a range of brightnesses.

\section{CONCLUSION}
\label{sec:conclusion}

The Hamamatsu OrcaQuest is an example of a new generation of sCMOS sensors -- CMOS sensors which boast many of the advantages of typical scientific CCD sensors such as low noise and high HDR, while also avoiding shortcomings related to the line-by-line readout systems and rolling shutters of CCD cameras. Since these cameras are a relatively new tool used for astronomical data collection, it is important to characterize these cameras properly. In the case of the OrcaQuest, we have identified shortcomings for any application that uses low exposure times -- such as lucky/speckle imaging -- or high reliance on HDR and relative signal brightnesses -- such as photometric  and spectroscopic analysis.

Although the OrcaQuest boasts very impressive linearity across its entire exposure range, demonstrating linearity that exceeds 99\% when using samples at linearly spaced intervals, this breaks down at low light levels. In the case of our detector, deviation from the detector's naive linear fit for datasets normalized ot 25 000 ADU/s exceeds 5\% at 2090 ADU and exceeds 10\% for signals under 356 ADU. This effect is mitigated through our calibration which reduces peak nonlinearity fivefold, and reduces the mean percentage residual from 1.27 to 1.02.

We find that this calibration is accurate between test runs with slight scattering up to 2\%, but identify little fixed relationship between brightness of the light source on the shape of the residual curve. Compared to a similar characterization experiment \cite{Strakhov_2023}, the peak residual is considerably higher at 1.6, versus another detector with 1.4, so it is recommended to recalibrate this curve according to each detector. For this detector, deviation of the residual curves between standard and UltraQuiet readout modes differ by up to 5\%, so it may or may not be recalibrated per readout mode depending on the accuracy needs of the sensor's application.

The method of data collection for this experiment is relatively simple and the calibration pipeline is publicly accessible, so we recommend doing so for any sensor utilizing the low exposure regime. In the future, care should be taken to reduce the scattering between datasets in order to speed up calibration time -- it is unclear whether this is related to temporal variations in the sensor, or issues with normalization of different sensor brightness levels.

\section{ACKNOWLEDGEMENTS}
\label{sec:acknowledgements}
The authors acknowledge support from the David A. Dunlap Department of Astronomy and Dunlap Institute for Astronomy and Astrophysics at the University of Toronto.  The Dunlap Institute is funded through an endowment established by the David Dunlap family and the University of Toronto.
This project was supported by the David A. Dunlap Seed Funding Program at the University of Toronto. In addition, this project is partially funded by the Canada Foundation for Innovation and Ontario Research Fund.

\newpage

\appendix    

\section{Additional plots for UltraQuiet readout}
\label{sec:uq}

   \begin{figure} [htbp]
   \begin{center}
   \includegraphics[height=8cm]{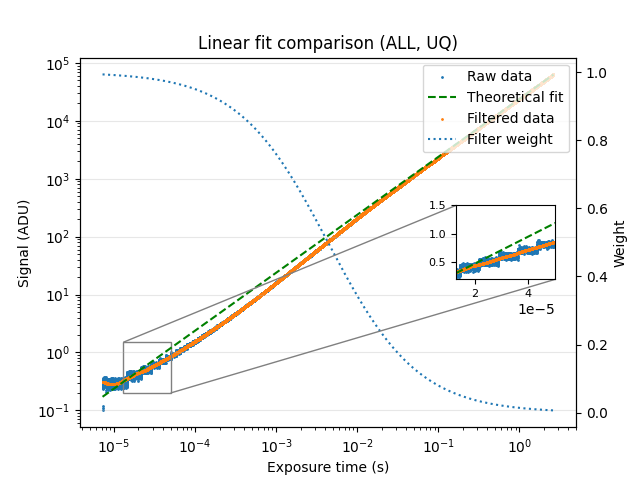}
	\end{center}
   \caption[example] 
   { \label{fig:linalluq} Linear fit, global signal levels by exposure, and smoothed data for UQ mode. See Figure \ref{fig:linallstd}.}.
   \end{figure}

   \begin{figure} [htbp]
   \begin{center}
   \includegraphics[height=8cm]{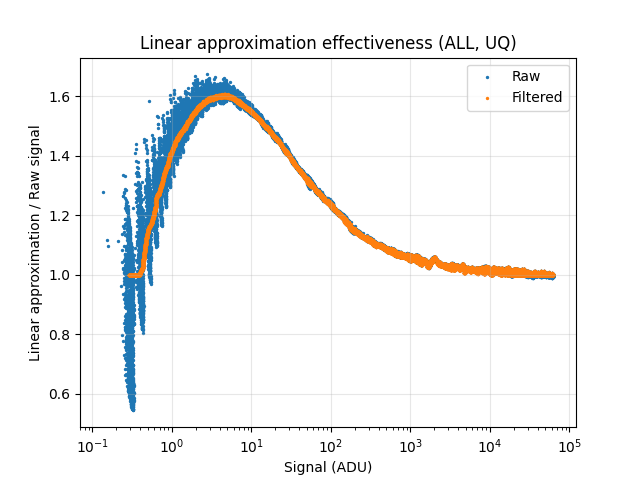}
	\end{center}
   \caption[example] 
   { \label{fig:rawscales2} Raw versus smoothed residuals for UQ readout mode. See Figure \ref{fig:rawscales}.}
   \end{figure} 
\newpage

\section{Suggested knot locations}
\label{sec:knots}
The initial knot locations used when modeling the B-spline calibration curve were $$[-0.42, -0.42, -0.42, -0.42, 2, 3.15, 3.15, 3.15, 3.15, 3.35, 3.5, 4.46, 4.46, 4.46, 4.46]$$
These represent the boundary values of ADU signal levels in log space for each polynomial interval.

The boundary values were further optimized using Nelder-Mead optimization to minimize the least-squares error over the entire signal range \cite{Gao2012Implementing}. An additional constraint was imposed to ensure the $3.15$ boundary remains within $0.05$ of its initial value in log space. This prevents the solver from passing over the 2 000 ADU bump by overfitting to a different location.

Although final values may vary from camera to camera, this is a useful starting point and may be used for all readout modes, pixels, and exposure brightnesses.

\bibliographystyle{plain} 
\bibliography{references}

\end{document}